\documentclass[pre,aps,twocolumn,showpacs,floatfix]{revtex4}

\usepackage[dvips]{graphicx}
\usepackage{amsmath}
\usepackage{amssymb}
\usepackage[nice]{nicefrac}
\usepackage{color}

\begin{document}

\title{Subordinated diffusion and CTRW asymptotics}

\author{Bart{\l}omiej Dybiec}
\email{bartek@th.if.uj.edu.pl}
\affiliation{Marian Smoluchowski Institute of Physics, and Mark Kac Center for Complex Systems Research, Jagellonian University, ul. Reymonta 4, 30--059 Krak\'ow, Poland}

\author{Ewa Gudowska-Nowak}
\email{gudowska@th.if.uj.edu.pl}
\affiliation{Marian Smoluchowski Institute of Physics, and Mark Kac Center for Complex Systems Research, Jagellonian University, ul. Reymonta 4, 30--059 Krak\'ow, Poland}

\date{\today}
\begin{abstract}
Anomalous transport is usually described either by models of continuous time random walks (CTRW) or, otherwise by fractional Fokker-Planck equations (FFPE).
The asymptotic relation between properly scaled CTRW and fractional diffusion process has been worked out via various approaches widely discussed in literature. Here, we focus on a correspondence between CTRWs and  time and space fractional diffusion equation stemming from  two different methods aimed to accurately approximate anomalous diffusion processes. One of them is the Monte Carlo simulation of uncoupled CTRW with a L\'evy $\alpha$-stable distribution of jumps in space  and a one-parameter Mittag-Leffler distribution of waiting times. The other is based on a discretized form of a subordinated Langevin equation in which the physical time defined via the number of subsequent steps of motion is itself a random variable. Both approaches are tested for their numerical performance and verified with known analytical solutions for the Green function of a space-time fractional diffusion equation.
The comparison demonstrates trade off  between precision of constructed solutions and computational costs. The method based on the subordinated Langevin equation leads to a higher accuracy of results, while the CTRW framework with a Mittag-Leffler distribution of waiting times provides efficiently an approximate fundamental solution to the FFPE and converges to the probability density function of the subordinated process in  a long-time limit.
\end{abstract}

\pacs{
 05.40.Fb, 
 05.10.Gg, 
 05.60.-k
 02.50.-r, 
 02.50.Ey, 
 }
\maketitle

%
%

\begin{quotation}
In the ubiquity of complex systems observed in Nature, non-Gaussian fluctuations prevail and transport properties deviate from the standard theory of Brownian motion.
Various facets of ``anomalous diffusion'' have been extensively studied over the past years by use of either continuous time random walks (CTRW) or fractional  kinetic equations. Recent development of simulation techniques based on Langevin equation in random subordinated time links uniquely stochastic trajectories of anomalous diffusion with probability density functions governed by fractional Fokker-Planck equations and provides efficient tools to study CTRW and their asymptotics. Our work quantifies convergence of CTRW to anomalous diffusion process
by using two schemes of modeling: We compare simulations of the Montroll-Weiss-Scher model with a Mittag-Leffler distribution of waiting times with subordinated Langevin dynamics. Efficiency of both methods in reproducing analytical results is tested along with analysis of numerical accuracy and computational costs.
\end{quotation}

\section{Introduction\label{sec:introduction}}

The continuous time random walk (CTRW) concept, as introduced in pioneering works by Montroll-Weiss-Scher \cite{klages2008} has been used over the decades for modeling diffusion processes on lattices, including anomalous transport. Commonly, the CTRW can be considered as a point process with reward \cite{cox1965}. Within such a framework evolution of a random walker position is described by a sequence of independent, identically distributed random positive variables $T_n$ which are interpreted as waiting times between consecutive jumps of the walker. In the simplest uncoupled version of the CTRW scenario, the jumps of length $x_n$ and waiting times are independent from each other. Correspondingly, the position of a diffusive particle can be represented by a random sum of independent random variables
\begin{equation}
X_{N(t)}\equiv \sum^{N(t)}_{n=1}x_n
\label{suma}
\end{equation}
with $N(t)$ staying for the counting random process which gives the (random) number of jumps up to a time $t$
\begin{equation}
N(t)=\mathrm{max}\left \{N: \sum^N_{n=1}T_n\leqslant t\right \}.
\end{equation}
A decoupled CTRW is Markovian only if the waiting time distribution $\psi(t)$ is exponential \cite{taylor1998}, i.e. when the corresponding counting process $N(t)$ is Poissonian. In this case, the CTRW process (\ref{suma}) has time-homogeneous (stationary) increments, i.e. it is an infinite divisible compound Poisson process.
So called dependent CTRW models (when the waiting times $T_n$ and the returns $x_n$ are coupled) were first studied by Shlesinger
et al. \cite{shlesinger1982} in order to place a physically realistic upper bound on particle velocities $x_n/T_n$.  Furthermore, as discussed by Meerschaert et al. \cite{meerschaert2006,meerschaert2004b} any coupled model at all for which in the long time scale ($c\rightarrow \infty$) the convergence $(c^{-1/\alpha}X_{[ct]},c^{-1/\nu}T_{[ct]})\Rightarrow (A(t),D(t))$ is met will have one of two kinds of limits: Either the dependence
disappears in the limit (because the waiting times and the jumps are asymptotically independent),
or else the limit process is one of those derivable from the Shlesinger's CTRW model \cite{shlesinger1982}. In such a case the counting process $N(t)$, as directly related to the jump-times process $T(n)$ by the relation
\begin{equation}
N(t)\geqslant n \Leftrightarrow T(n)\leqslant t,
\label{steps}
\end{equation}
follows the inverse random time distribution \cite{piryatinska2005,saichev1997,uchaikin2003} with the waiting time defined as the inter-jump time interval $T(n)-T(n-1)$. Put it differently, both processes $T(n)$ and $N(t)$ can be viewed as mutually inverse random functions leading to the equivalence
\begin{eqnarray}
\mathrm{Prob}\left \{T(n) < t \right \} & = & \mathrm{Prob}\left \{N(t)\geqslant n \right \} \nonumber \\
& = & \int^{\infty}_np_{n'}(n'(t))dn'.
\end{eqnarray}
In the limit when $n$ becomes a continuous parameter \cite{piryatinska2005,saichev1997,uchaikin2003}, for which the variable $T(n)$ is assumed to be distributed according to a strictly asymmetric, one sided $\nu$-stable distribution $L_{\nu,1}$, the probability density (PDF) $p_{n}(n(t))$ can be obtained from  the corresponding PDF of the random time $T(n)$
\begin{equation}
p_{n}(n(t))=-\frac{\partial}{\partial n}\int^{t}_{0}l_{\nu,1}(t';n)dt'
\label{inverse}
\end{equation}
where $l_{\nu,1}(t;n) =dL_{\nu,1}(t;n)/dt$.
Accordingly, the longtime
limit process $X_{N(t)}$ is then described by the probability density
 \cite{tunaley1974,saichev1997,shlesinger1995,kotulski1995,nielsen2001,barkai2002,uchaikin2003,srokowski2009},
\begin{equation}
p(x,t)=\int^{\infty}_0 p(x,\tau)p_n(\tau,t) d\tau
\end{equation}
which possesses the scaling property
\begin{equation}
 p(x,t)=t^{-\nu/\alpha}p(xt^{-\nu/\alpha},1).
\label{eq:scaling}
\end{equation}
The time series Eq.~(\ref{suma}) can be otherwise characterized by the probability of jumps $x(t_{n+1})-x(t_n)$ and waiting times $t_{n+1}-t_n$. In case that these both variables are statistically independent with
the waiting time  $\psi(t)$ and the jump length $p(x)$ PDFs having tails ($x\rightarrow\infty$, $t\rightarrow\infty$) of the power-law type
\begin{equation}
\psi(t) \propto t^{-(\nu+1)},
\label{eq:wtime}
\end{equation}
and
\begin{equation}
p(x) \propto |x|^{-(\alpha+1)},
\label{eq:jlength}
\end{equation}
the Laplace-Fourier transform $p(q,u)$  of the PDF $p(x,t)$ takes the form \cite{shlesinger1995,saichev1997}
\begin{eqnarray}
p(q,u) & = & \frac{u^{\nu-1}}{u^{\nu}+|q|^{\alpha}} \nonumber \\
& = & u^{\nu-1}\int^{\infty}_0 ds \exp\left[-s(u^{\nu}+|q|^{\alpha})\right].
\end{eqnarray}
Here $0<\nu < 1$ and $0<\alpha < 2$ are stability indices of the corresponding time and jump length distributions.

The CTRW scheme as described by Eqs.~(\ref{eq:wtime}) and (\ref{eq:jlength}) asymptotically leads to the situation when the evolution of the probability density $p(x,t)$ of finding a random walker at  the position $x$ after $n$ steps performed up to time $t$ can be described by a  continuous integral $p(x,t)=\int^{\infty}_0d\tau p(x,\tau)p_n(\tau,t)$ being the solution to the fractional Fokker-Planck equation \cite{saichev1997,metzler1999,barkai2002}
\begin{equation}
 \frac{\partial p(x,t)}{\partial t}={}_{0}D^{1-\nu}_{t}\left[  \frac{\partial^\alpha p(x,t)}{\partial |x|^\alpha} \right] ,
\label{eq:ffpe}
\end{equation}
with the initial condition $p(x,0)=\delta(x)$. In the above equation ${}_{0}D^{1-\nu}_{t}$ denotes the Riemann-Liouville fractional (time) derivative ${}_{0}D^{1-\nu}_{t}=\frac{d}{d t}{}_{0}D^{-\nu}_{t}$ defined by the relation
\begin{equation}
{}_{0}D^{1-\nu}_{t}f(x,t)=\frac{1}{\Gamma(\nu)}\frac{d}{d t}\int^{t}_0 dt'\frac{f(x,t')}{(t-t')^{1-\nu}}
\end{equation}
and $\frac{\partial^{\alpha}}{\partial |x|^{\alpha}}$ stands for the Riesz-Weyl fractional (space) derivative with the Fourier transform ${\cal{F}}[\frac{\partial^{\alpha} f(x)}{\partial |x|^{\alpha}} ]=-|k|^{\alpha}\mathcal{F}\left[ f(x) \right]$.
In the limit of $\alpha=2$ and $\nu=1$, a L\'evy random walk (\ref{suma}) is (asymptotically) equivalent to the standard (Markovian) Brownian motion, while for $\nu=1$ with $\alpha<2$ it corresponds to (Markovian) L\'evy flights.

Both approaches are in use and have received attention and application in a plethora of physical
and biological problems such as mass and charge transport in disordered systems, relaxation phenomena, front propagation in reaction-diffusion systems, transport in plasma, motion of organelles or epidemic spread \cite{sokolov2000,barkai2000,chechkin2002c,meerschaert2004,gudowska2005,szabat2007,Heinsalu2007,koren2007c,klages2008,sokolov2009,gudowskanowak2009b}.
Suitability of the fractional dynamics by means of the coupled Langevin equations representing the subordination scheme has been also proved in analysis of correlation functions \cite{baule2007}.

Our paper is devoted to a detailed analysis of the CTRW diffusive limit obtained based on two different approaches. First relates to the role played by Mittag-Leffler functions in anomalous relaxation \cite{klages2008,hilfer1995}. In particular, we examine the uncoupled case of the CTRW scenario which includes implicitly the Mittag-Leffler waiting time distribution function
\begin{equation}
E_{\nu}(z)=\sum^{\infty}_{n=0}\frac{z^n}{\Gamma(\nu n+1)}
\end{equation}
being solution of the equation
\begin{equation}
{}_{0}D^{\nu}_{t} \Psi(\tau)=-\Psi(\tau)
\end{equation}
for the so called survival function $\Psi(\tau)\equiv1-\int^{\tau}_0\psi(s)ds$, $\Psi(\tau)=E_{\nu}(-\tau^{\nu})$.
The second choice is based on the CTRW modeling involving a subordination technique \cite{fogedby1994,eule2009,magdziarz2007}.

The paper is organized as follows:
In Section~\ref{sec:relation} some generic properties of both approaches are introduced. Section~\ref{sec:results} presents results of numerical analysis, in which asymptotic properties of the subordinated Langevin equation are compared with the asymptotics of the CTRW scheme with $\psi(t)$ given by a one-parameter Mittag-Leffler distribution and $p(x)$ represented by a symmetric L\'evy $\alpha$-stable distribution. The supremacy of subordination technique is further demonstrated in Section~\ref{sec:numissues} which discusses numerical costs of computation and precision of both methods applied. The paper is closed with concluding remarks.

%
%
\section{Relation between CTRW, subordination and fractional calculus\label{sec:relation}}

In the limited number of cases Eq.~(\ref{eq:ffpe}) can be solved analytically \cite{saichev1997,metzler2002}. Numerical methods of solving the fractional Fokker-Planck equation (\ref{eq:ffpe}) resume usually two directions. First, it is possible to approximate the solving probability density $p(x,t)$. This can be achieved by numerical approximation to the fractional derivatives present in Eq.~(\ref{eq:ffpe}), see \cite{podlubny1999,meerschaert2004,Chen2004}. Nevertheless, due to a nonlocal character of fractional derivatives, the convergence of approximation schemes can be very slow. The other possibility relies on the construction of the ensemble of trajectories from which the estimators of $p(x,t)$ can be derived in the form of the frequency histograms. Realizations of the stochastic process, whose probability density evolves according to Eq.~(\ref{eq:ffpe}), can be constructed either by the subordination technique \cite{eule2009,magdziarz2007,gorenflo2007,mura2008} or by the CTRW framework \cite{scalas2006,fulger2008}.  Here we compare these two approaches \cite{magdziarz2007,fulger2008} and discuss their rate of convergence to known, analytical solutions of corresponding fractional Fokker-Planck equations.

In a series of papers \cite{magdziarz2007b,magdziarz2008,magdziarz2007} subordination methods \cite{feller1968,eule2009,piryatinska2005} have been extended to give a stochastic representation of trajectories of the process $X(t)$ which is otherwise described by the fractional Fokker-Planck equation (\ref{eq:ffpe}). Within the subordination approach the process of primary interest $X(t)$ is obtained as a function $X(t)=\tilde{X}(S_{\nu}(t))$ by randomizing the time ``clock'' of the process $\tilde{X}(s)$ using a different ``clock'' $S_{\nu}(t)$ which links the real time $t$ with the operational time $s$. Accordingly, the stochastic representation of the solution to Eq.~(\ref{eq:ffpe}) is obtained in this scheme by use of the self similar process $S_{\nu}(t)$ representing so called $\nu$-stable (inverse) subordinator. The latter is the process with nonnegative increments \cite{piryatinska2005} and can be defined as
\begin{equation}
S_{\nu}(t)=\mathrm{inf}\left \{s>0:T(s)>t \right \}.
\label{eq:subordinator}
\end{equation}
In the above equation $T(s)$ denotes a strictly increasing $\nu$-stable process ($0<\nu<1$) whose Laplace transform is given by $\langle e^{-kT(s)} \rangle =e^{-sk^{\nu}}$ whereas its inverse $S_{\nu}(t)$ determines random hitting time (first passage time) for the problem \cite{meerschaert2004b,piryatinska2005}.
 In turn, the parent process $\tilde{X}(s)$ is composed of increments of symmetric $\alpha$-stable motion
\footnote{Recall that the $\alpha$-stable motion $\left \{L_{\alpha}(t), t\geqslant 0\right \}$ is defined \cite{janicki1994} as a stochastic process of independent, stationary increments $\Delta L_{\alpha}=L_{\alpha}(t)-L_{\alpha}(t')$ distributed according to a stable law of index $\alpha$.}
described in an operational time $s$ by the equation
\begin{equation}
d\tilde{X}(s)=dL_{\alpha,0}(s).
\label{eq:langevineq}
\end{equation}
The combination of Eqs.~(\ref{eq:subordinator}) and (\ref{eq:langevineq}) fully determines the process $X(t)=\tilde{X}(S_{\nu}(t))$ and provides a stochastic representation to Eq.~(\ref{eq:ffpe}) in terms of the ensemble of trajectories \cite{magdziarz2007b,magdziarz2008,magdziarz2007,dybiec-anomalous}.

In a less formal, albeit quite intuitive  way, description of  continuous realization of the CTRW scheme Eqs.~(\ref{suma})--(\ref{inverse})
has been proposed by Fogedby \cite{fogedby1994} and Eule \cite{eule2009}. Their formulation of subordination procedure is based on the analysis of the set of coupled Langevin equations
\begin{equation}
\left\{
\begin{array}{l}
\dot{x}(s)\equiv \frac{dx}{ds}=\xi(s) \\
\dot{t}(s)=\eta(s) \\
\end{array}
\right.,
\label{complex}
\end{equation}
where the random walk $x(t)$ becomes parametrized by variable $s$. In the above equations $\xi(s)$ and $\eta(s)$ are assumed to be independent, random noises and the pair process $(x(s),t(s))$ preserves the Markov property. The requirement of causality ($t(s)$ is a physical time) limits choice of $\eta(s)$ to functions returning positive values only. The combined process in physical time $t$ is described by the trajectories $x(t)=x(s(t))$ and is  subordinated to the parent process with corresponding realizations $x(s)$. Moreover, the time transformation implies
\begin{equation}
s(\tau)=T\Leftrightarrow t(T)=\tau,
\end{equation}
or
\begin{equation}
\mathrm{Prob}\left \{s(\tau)<T \right \}=\mathrm{Prob}\left \{t(T)\geqslant \tau \right \}.
\end{equation}
Let $\xi(t)$ stands for a white, symmetric L\'evy noise \cite{dybiec2006,dubkov2008}. In this case, the increments $\Delta x(s)$ are assumed independent over non-overlapping time intervals, i.e. they define  a stationary process being a generalization of a Brownian motion (a generalized Wiener process). Accordingly, the solution of the stochastic differential equation $\dot{x}(s)=\xi(s)$ can be expressed in the form ($N\Delta s' = s$)
\begin{equation}
 x(s)=L_{\alpha,0}(s)=\int^s_0ds'\xi(s') \approx \sum^{N-1}_{i=0}(\Delta s')^{1/\alpha}\xi_i
\end{equation}
 with the PDF $l_{\alpha,0}(x)$ whose Fourier transform $\Phi(k,s)=\left \langle e^{ikx(s)}\right \rangle=\left \langle e^{ikL_{\alpha,0}(s)}\right \rangle$ is given by
\begin{equation}
\Phi(k,s) = \exp\left[ -s|k|^\alpha \right].
\label{eq:charakt}
\end{equation}
In turn, let us assume that the stationary random process $\eta(s)$ is a white $\nu$-stable L\'evy noise which takes positive values only. In consequence, the integrated process
\begin{equation}
t(s)=\int^s_0ds'\eta(s')
\end{equation}
is a $\nu$-stable totally skewed L\'evy motion with an index $0<\nu<1$ and characteristic function given by
\begin{equation}
\Phi(k,s) = \exp\left[ -s|k|^\nu  \left (1-i\mathrm{sign} k\tan\frac{\pi\nu}{2} \right)\right].
\label{time}
\end{equation}
 The PDF of the random variable $s$ at time $t$, $p(s,t)$  is then given by the inverse stable distribution
 \begin{eqnarray}
p(s,t) & = & -\frac{d}{ds}\int_0^t \Theta(y,s) dy \nonumber \\
& = & \frac{d}{ds}(1-L_{\nu,1}(t/s^{1/\nu})),
\end{eqnarray}
with the process $s(t)$ being an asymptotic (continuous) analog of the number of steps $n(t)$, cf. Eqs. (\ref{steps})--(\ref{inverse}).
In the above $\Theta(y,s)$ stands for the PDF of times $t$ with the Fourier transform $\left \langle e^{-kt(s)}\right \rangle $ given by Eq.~(\ref{time}). Again, the PDF $p(x,t)$ of the subordinated process $x(t)$ coincides \cite{fogedby1994,magdziarz2007} with the solution to the fractional Fokker-Planck equation (\ref{eq:ffpe}).

Let us stress that the numerical methods to approximate the solution of the Langevin equations (\ref{complex}) involve an Ito integral with respect to the (generalized) Brownian motion and assume the discretization of the time parameter. The computer algorithm generates then the approximation of the trajectory (a single realization of the process $X(t)$) in terms of a random walk on a one-dimensional grid, i.e. it simulates sample paths of a corresponding CTRW. Therefore both, subordination and CTRW methods, are essentially different facets of the random walk methodology. However, the method based on the subordinated Langevin equation provides exact representation of solutions of the fractional Fokker-Planck equation, while CTRW scenarios reconstructs solutions asymptotically.

The fractional time derivative in Eq.~(\ref{eq:ffpe}) results in the Mittag-Leffler decay of temporal eigensolutions.
Consequently, the alternative approximation \cite{fulger2008} to Eq.~(\ref{eq:ffpe}) relies on the generation of waiting times distributed according to the complementary distribution function given by the Mittag-Leffler function \cite{kozubowski1999} while the jump lengths are distributed according to the $\alpha$-stable density.

The probability density $p(x,t)$, see Eq.~(\ref{eq:ffpe}), is known to have a series representation \cite{saichev1997}:
\begin{eqnarray}
p(x,t) & = & \frac{1}{\pi|y|t^{\nu/\alpha}}\sum^{\infty}_{k=0}\frac{(-1)^k}{|y|^{k\alpha}} \frac{\Gamma(k\alpha+1)}{\Gamma(k\nu+1)}\nonumber \\
& & \times \cos\left[\frac{\pi}{2}(k\alpha+1)\right]
\label{eq:series}
\end{eqnarray}
where $y=x/t^{\nu/\alpha}$. The above series are divergent for $\alpha\geqslant\nu$. However, for $\alpha=\nu$, the summation has been shown \cite{saichev1997} to produce the closed analytical formula
\begin{equation}
p(x,t)=\frac{1}{\pi |y|t}\frac{\sin(\pi\nu/2)}{|y|^{\nu}+|y|^{-\nu}+2\cos(\pi\nu/2)}
\label{eq:closedformula}
\end{equation}
where (as previously) $y=x/t^{\nu / \alpha}$.
In the limit of $\alpha=2$ and $\nu=1$ the standard Markovian (Brownian) diffusion is recovered.

Exact solutions to Eq.~(\ref{eq:ffpe}) are known in special cases \cite{saichev1997,metzler2002}:
For $\nu=1$ with any value of $\alpha$, $p(x,t)$ is an $\alpha$-stable process
\begin{equation}
p(x,t)=l_{\alpha}( t^{1/\alpha}x) .
\end{equation}
In particular for $\alpha=2$ the Gaussian distribution is obtained
\begin{equation}
p(x,t)= \frac{1}{\sqrt{4\pi t}} \exp\left[ -\frac{x^2}{4t}\right],
\label{eq:gauss}
\end{equation}
while for $\alpha=1$ the Cauchy distribution is reached
\begin{equation}
p(x,t)=\frac{t}{\pi } \frac{1}{x^2+t^2}.
\label{eq:cauchy}
\end{equation}
Analytical solutions are also known for $\nu=1/2,\;\alpha=1$ \cite{saichev1997}
\begin{equation}
p(x,t)=-\frac{1}{2\pi^{3/2}\sqrt{t}}\exp\left[\frac{x^2}{4t}\right]\mathrm{Ei}\left[-\frac{x^2}{4t}\right],
\label{eq:pa1n12}
\end{equation}
and for $\nu=2/3,\;\alpha=2$ (see \cite{saichev1997})
\begin{equation}
p(x,t)=\frac{3^{2/3}}{2t^{1/3}}\mathrm{Ai}\left[ \frac{|x|}{(3t)^{1/3}} \right],
 \label{eq:pa2n23}
\end{equation}
where $\mathrm{Ei}(x)$ and $\mathrm{Ai}(x)$ are the integral exponential function and the Airy function respectively.

\begin{figure}[!ht]
\begin{center}
\includegraphics[angle=0,width=8.0cm]{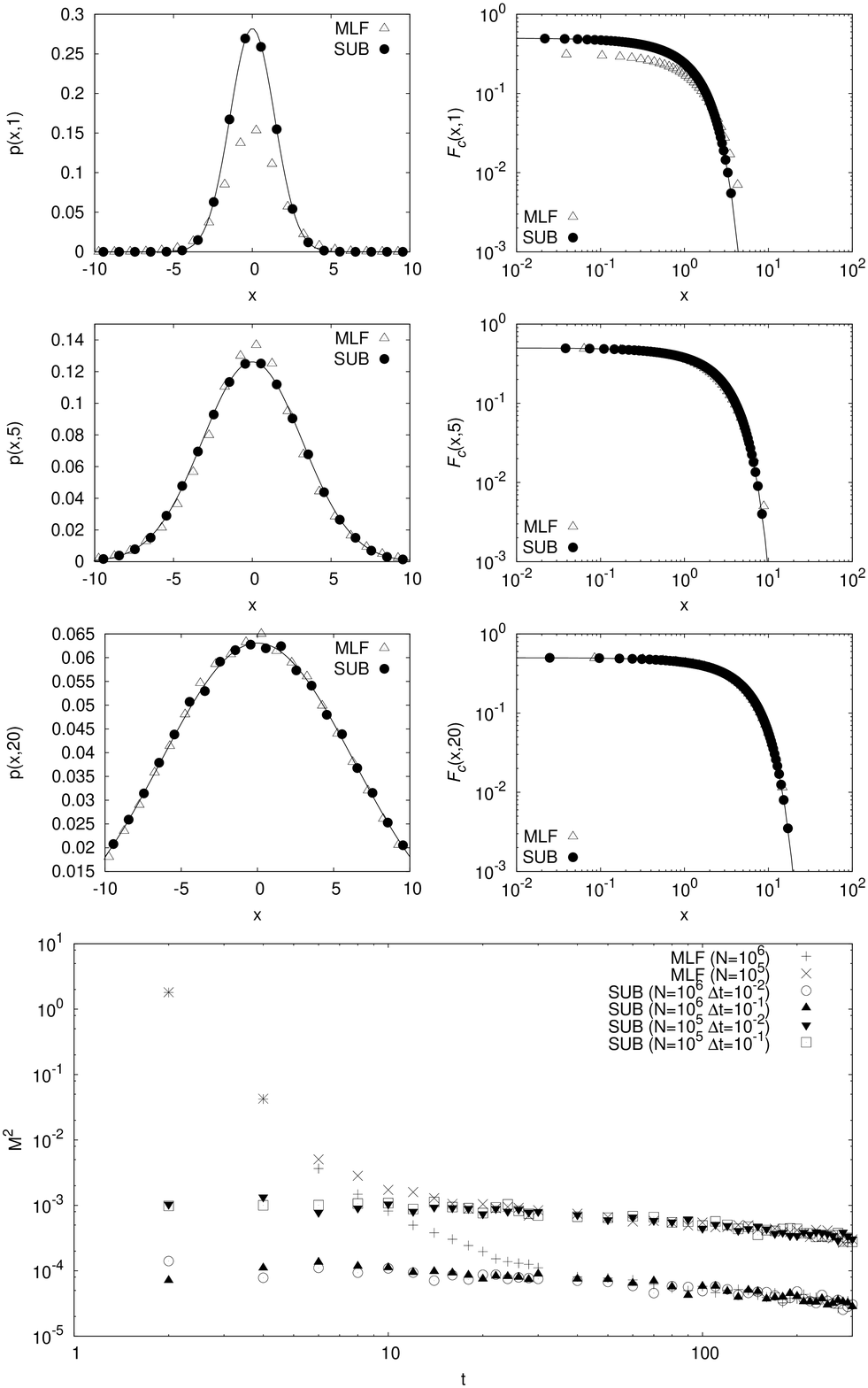}
\caption{Probability density functions $p(x,t)$ (left panel) and complementary cumulative density functions $F_c(x)=1-F(x)$  (right panel) for the subdiffusion parameter $\nu=1.0$ and the stability index $\alpha=2.0$. Results were constructed using the subordination method (black circles) and the CTRW framework with the Mittag-Leffler waiting time distribution (empty triangles). Thin solid black line represents analytical solution given by Eqs.~(\ref{eq:gauss}) and (\ref{eq:hdf}). Various panels correspond to various times $t=\{1,5,20\}$ (from top to bottom). Finally, bottom panel presents the sum of squared differences, $M^2$, between analytical and numerical results (see Eq.~(\ref{eq:m2})) for $\nu=1.0$ $\alpha=2.0$. Various curves compare simulation results constructed by the subordination and CTRW methods with various time steps of the integration ($\Delta t$) and number of repetitions ($N$).}
\label{fig:s10a20}
\end{center}
\end{figure}

%
%
\section{Results\label{sec:results}}

\begin{figure}[!ht]
\begin{center}
\includegraphics[angle=0,width=8.0cm]{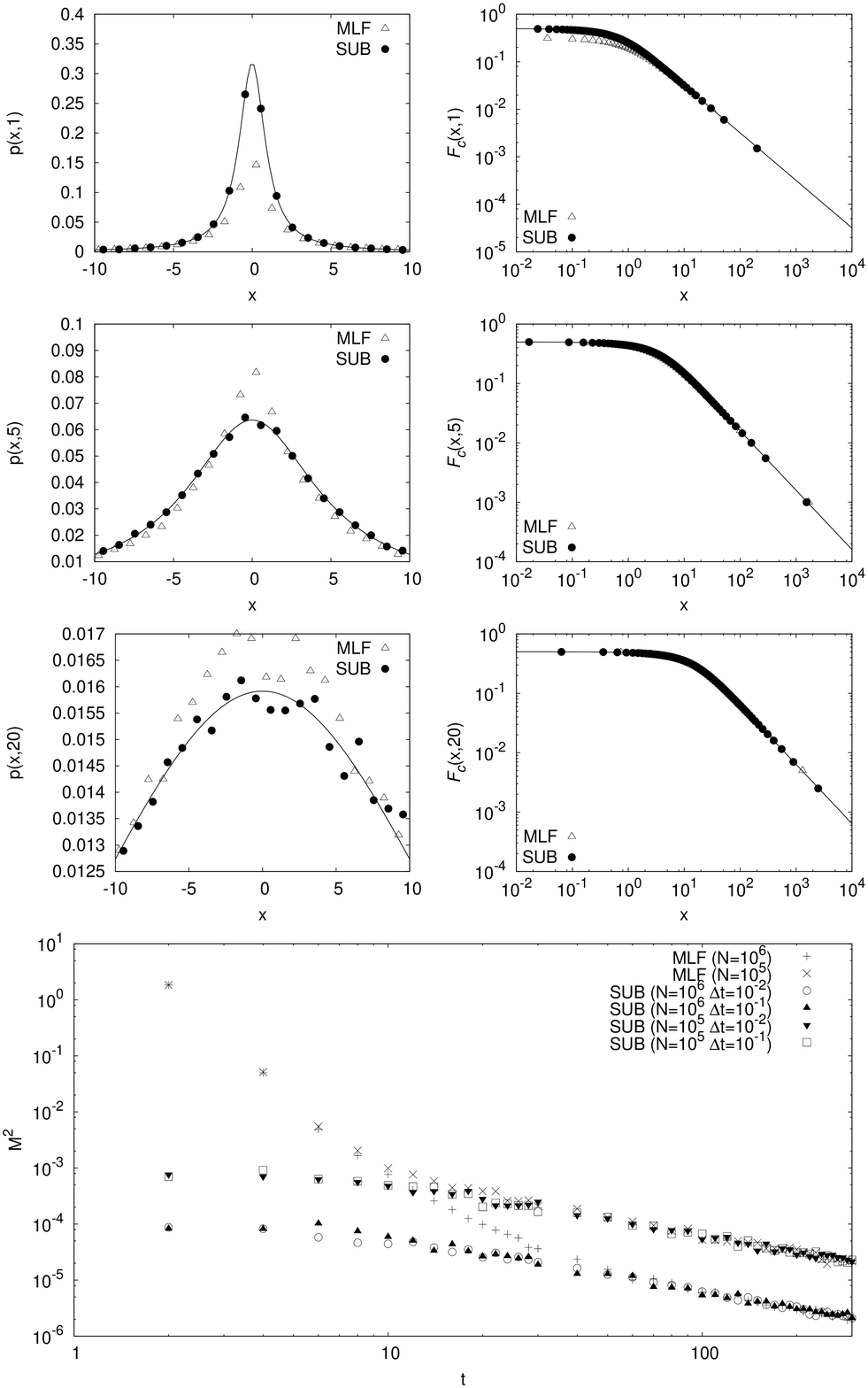}
\caption{The same as in Fig.~\ref{fig:s10a20} for the subdiffusion parameter $\nu=1.0$ and the stability index $\alpha=1.0$ and Eq.~(\ref{eq:cauchy}).}
\label{fig:s10a10}
\end{center}
\end{figure}

\begin{figure}[!ht]
\begin{center}
\includegraphics[angle=0,width=8.0cm]{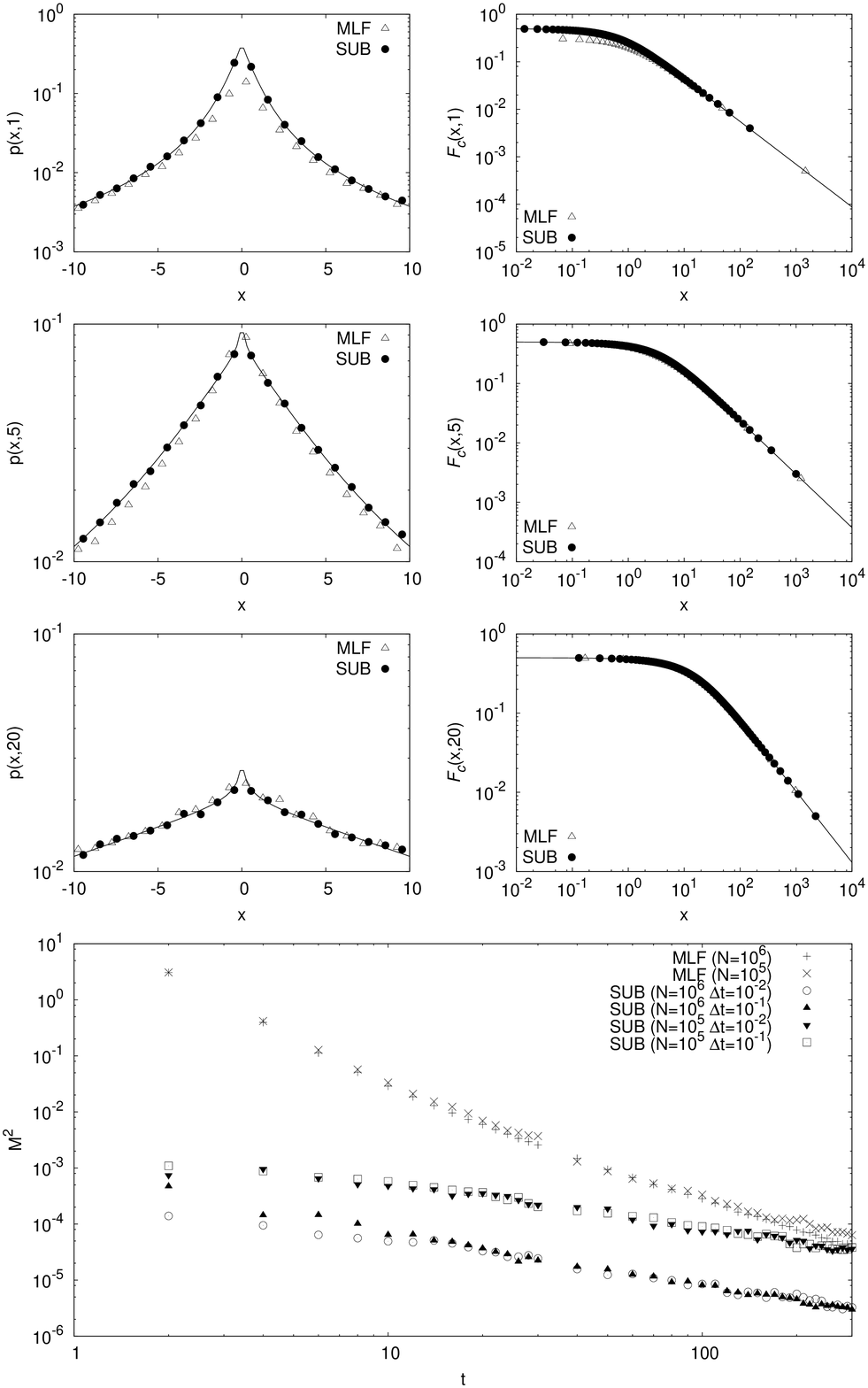}
\caption{The same as in Fig.~\ref{fig:s10a20} for the subdiffusion parameter $\nu=0.9$ and the stability index $\alpha=0.9$ and Eq.~(\ref{eq:closedformula}).}
\label{fig:s09a09}
\end{center}
\end{figure}

\begin{figure}[!ht]
\begin{center}
\includegraphics[angle=0,width=8.0cm]{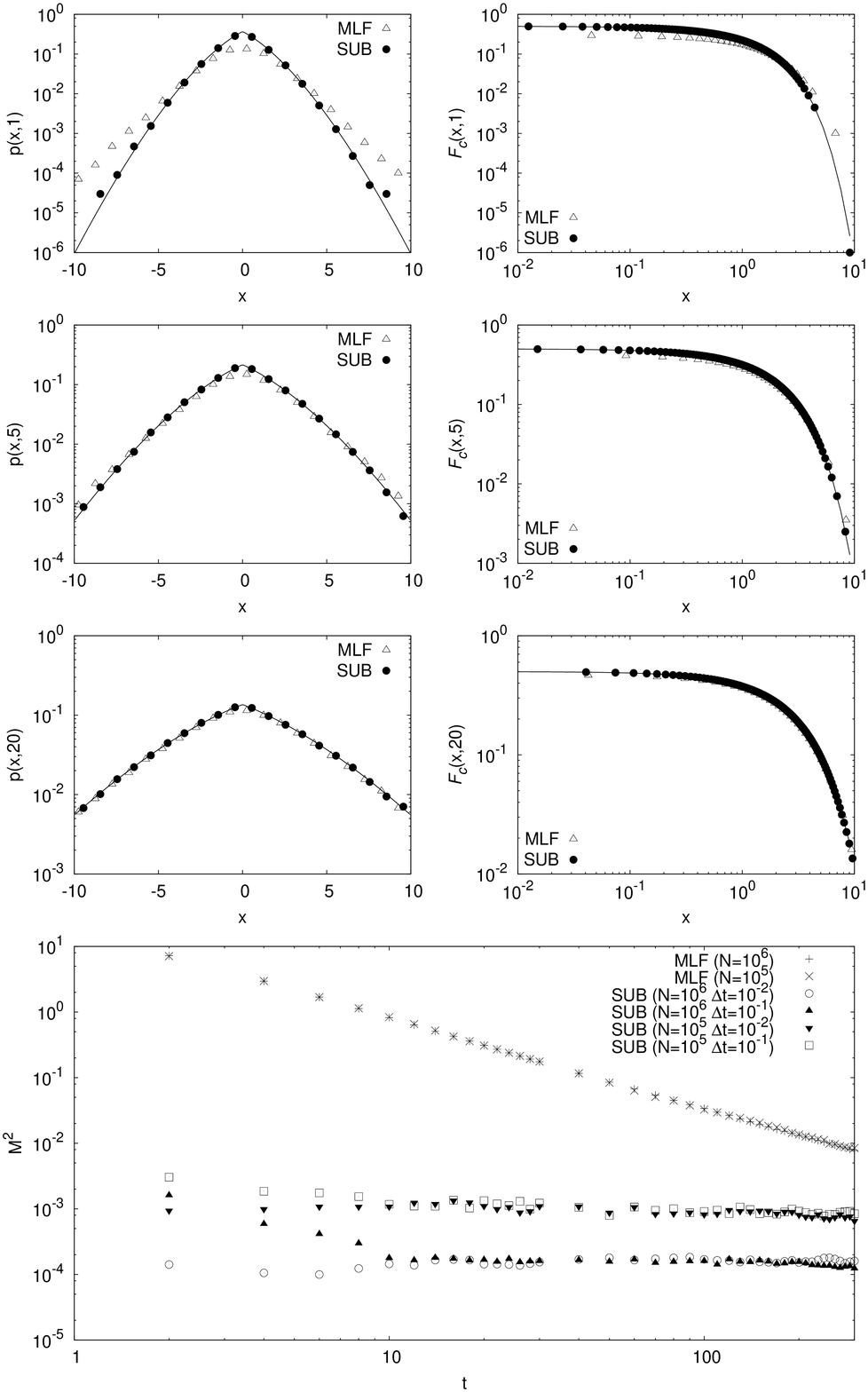}
\caption{The same as in Fig.~\ref{fig:s10a20} for the subdiffusion parameter $\nu=2/3$ and the stability index $\alpha=2.0$ and Eq.~(\ref{eq:pa2n23}).}
\label{fig:s066a20}
\end{center}
\end{figure}

\begin{figure}[!ht]
\begin{center}
\includegraphics[angle=0,width=8.0cm]{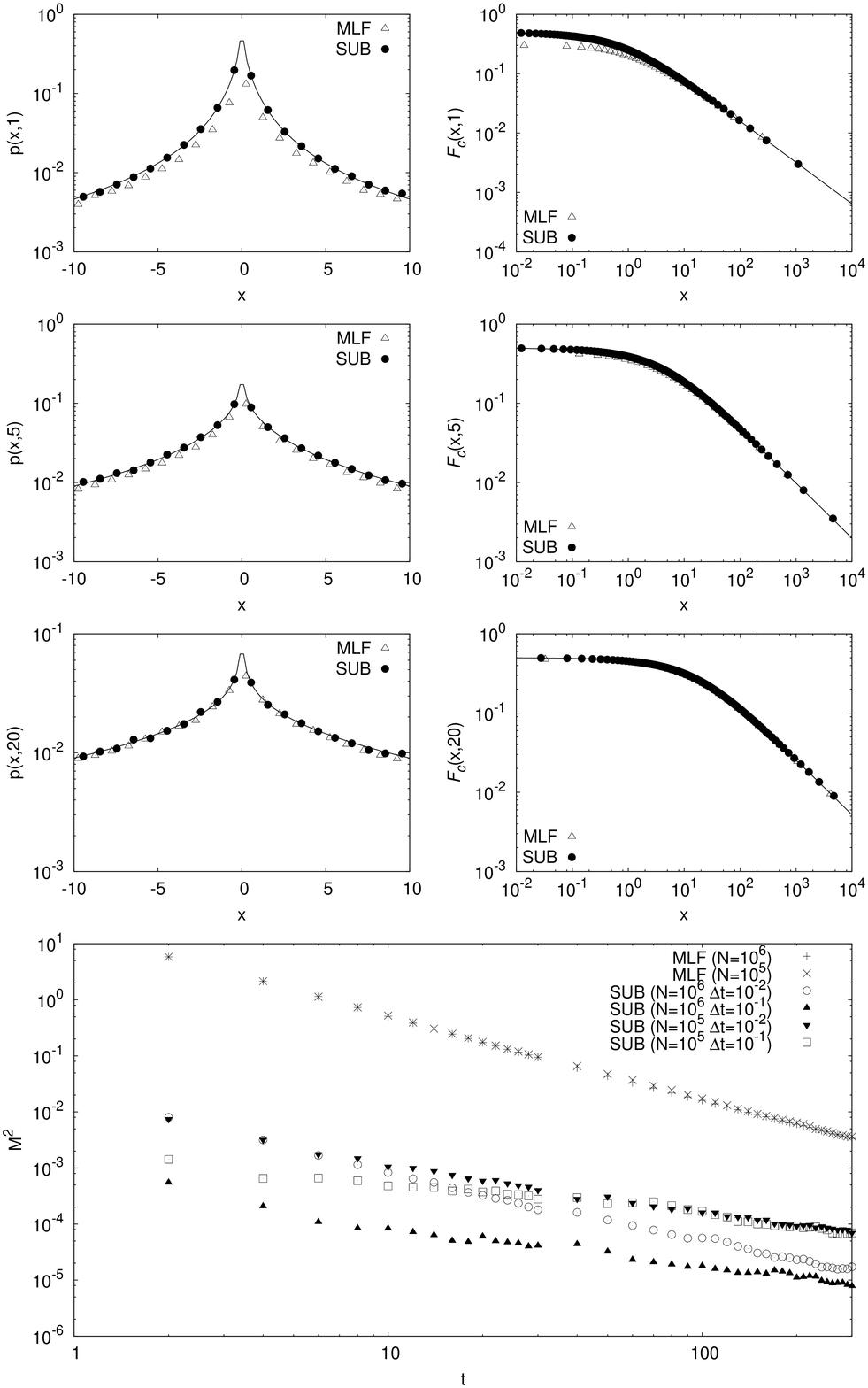}
\caption{The same as in Fig.~\ref{fig:s10a20} for the subdiffusion parameter $\nu=0.7$ and the stability index $\alpha=0.7$ and Eq.~(\ref{eq:closedformula}).}
\label{fig:s07a07}
\end{center}
\end{figure}

\begin{figure}[!ht]
\begin{center}
\includegraphics[angle=0,width=8.0cm]{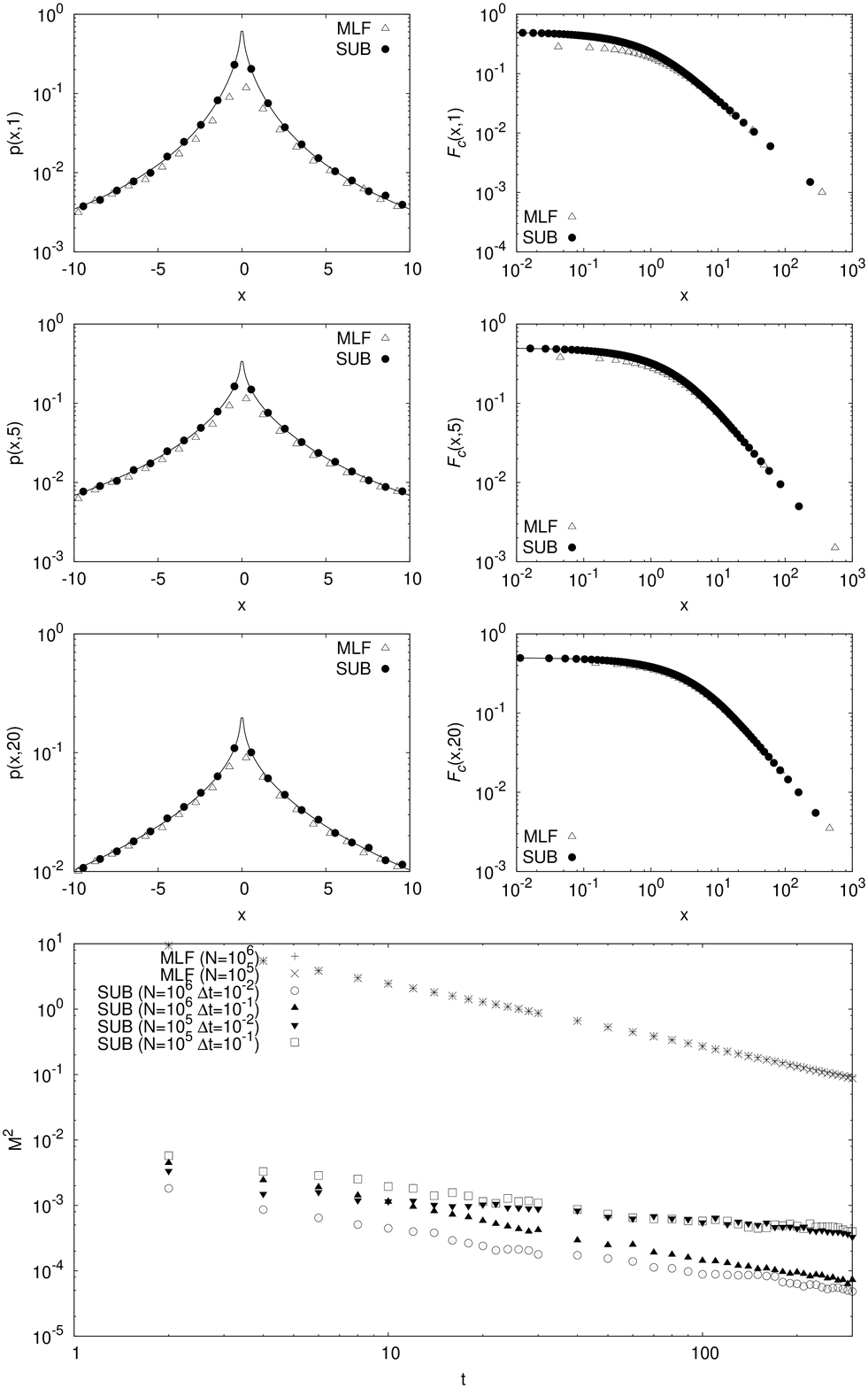}
\caption{The same as in Fig.~\ref{fig:s10a20} for the subdiffusion parameter $\nu=0.5$ and the stability index $\alpha=1.0$ and Eq.~(\ref{eq:pa1n12}).}
\label{fig:s05a10}
\end{center}
\end{figure}

Left panels of Figs.~\ref{fig:s10a20} -- \ref{fig:s05a10} present sample probability density functions $p(x,t)$ estimated at various times $t=\{1,5,20\}$. For each parameter set, numerical results were constructed separately by the subordination technique \cite{magdziarz2007}  and  the CTRW method \cite{fulger2008}. Furthermore, to ascertain the correctness of both methods, verification of obtained solutions to the FFPE has been  examined against analytical formula for several known cases  Eqs.~(\ref{eq:closedformula}), (\ref{eq:gauss}), (\ref{eq:cauchy}), (\ref{eq:pa1n12}) and (\ref{eq:pa2n23}). Derived results were averaged over $N=10^6$ realizations of the corresponding stochastic process. In the subordination method the time step of the integration was adjusted to $\Delta t=10^{-2}$. Examples of the histograms generated according to both procedures, see  Figs.~\ref{fig:s10a20} -- \ref{fig:s05a10}, demonstrate that the subordination method not only reconstructs properly the tails of probability densities but also matches correctly their central parts.

The comparison between numerical solutions and their analytical analogues (see left panels of Figs.~\ref{fig:s10a20} --  \ref{fig:s05a10}) is based on the analysis of the  sum of squared deviations:
\begin{equation}
 M^2=\sum_{i=1}^{N_{b}}[p(x_i,t)-\hat{p}(x_i,t)]^2.
\label{eq:m2}
\end{equation}
In Eq.~(\ref{eq:m2}), $N_b$ represents number of histograms bins, $x_i$ locations of bin centers, $p(x,t)$ and $\hat{p}(x,t)$ denote analytical and estimated probability density functions, respectively. Various curves in bottom panels of Figs.~\ref{fig:s10a20} -- \ref{fig:s05a10} correspond to different method of construction of solutions or different simulation parameters.

In general, results based on the subordination technique reconstruct much better the shape of theoretical probability density. For the Markovian case $\nu=1$, and after sufficiently long time, similar accuracy of both methods is observed, see bottom panels of Figs.~\ref{fig:s10a20} and \ref{fig:s10a10}. Furthermore, in the Markovian case the level of agreement between analytical and numerical results depends only on the number of repetitions, i.e. both the subordination and CTRW methods have similar accuracy, see bottom panels of Figs.~\ref{fig:s10a20} -- \ref{fig:s10a10}.  On the contrary to the Markovian ($\nu=1$) case, for  a non-Markov process ($\nu<1$),  differences between both methods are well visible even for long times, see Figs.~\ref{fig:s09a09}, \ref{fig:s066a20}, \ref{fig:s07a07} and \ref{fig:s05a10}. Here, the level of agreement depends both on the method applied and the number of repetitions. The method based on the CTRW with the one-parameter Mittag-Leffler distribution of waiting times (with $10^5$ or $10^6$ repetitions) leads to the same level of agreement. In contrast, the approach based on the subordination results in significantly smaller deviations from theoretical distributions. Also, expanding the ensemble of simulated trajectories (by  increasing the number of repetitions) clearly increases the level of agreement between analytical and numerical solutions. In the force free case, which is studied here, the choice of the times step of integration seems to be less important than the choice of the number of repetitions. Finally, the increase in the histogram range (with fixed number of bins) leads to smaller values of $M^2$ because spatial resolution of the histogram is decreased (results not shown). Nevertheless, such a comparison still demonstrates better performance of the subordination method than the CTRW framework.

Right panels of Figs.~\ref{fig:s10a20} -- \ref{fig:s05a10} present sample complementary cumulative distributions $F_c(x,t)$ for various times $t=\{1,5,20\}$
\begin{equation}
 F_c(x,t)=1-F(x,t)=1-\int_{-\infty}^xp(x',t)dx'
\label{eq:hdf}
\end{equation}
with their analytical counterparts obtained after integration of corresponding PDFs $p(x,t)$  given by Eqs.~(\ref{eq:closedformula}), (\ref{eq:gauss}), (\ref{eq:cauchy}), (\ref{eq:pa1n12}) and (\ref{eq:pa2n23}).
 Right panels of Figs.~\ref{fig:s10a20} -- \ref{fig:s05a10} demonstrate that both methods perfectly reconstruct the asymptotic dependence of the probability densities.

The properties of solutions to the FFPE (\ref{eq:ffpe}) are determined by the subdiffusion parameter $\nu$ and the stability index $\alpha$.
In the simulations based on a subordination scheme, the subordinator process $S_{\nu}(t)$ is evaluated by generating first increments of the process $T(s)$, cf. Eq.~(\ref{eq:subordinator}), for which $S_{\nu}(t)$ forms the inverse. Since $T(s)$ is a L\'evy jump process with nonnegative increments, every jump of $T(s)$ can be associated with a long waiting time of its inverse $S_{\nu}(t)$ \cite{piryatinska2005,magdziarz2007c}. This heavy-tailed distribution of waiting times is a feature of subdiffusive dynamics and is responsible for a weak ergodicity breaking and a non-Markov character of the combined process $X(t)$  \cite{rebenshtok2007,lomholt2007,he2008,rebenshtok2008,lubelski2008b,dybiec2009b,dybiec2009h}.

The power law distribution of waiting times (for $\nu<1$ the mean waiting time is divergent) is also properly reconstructed by an explicit use of the Mittag-Leffler distribution of the jump-times in the CTRW  scheme.
In fact, the Mittag-Leffler distribution interpolates between  the short-time stretched-exponential distribution of waiting times and the long time power-law asymptotics \cite{metzler2000}. The resulting long-range memory of simulated PDFs is well visible in histograms where (for $\nu<1$) a persistent cusp at $x=0$ is detected (see left panels of Figs.~\ref{fig:s09a09} -- \ref{fig:s05a10}). This behavior typical for subdiffusive systems \cite{sokolov2002} is also manifested in ambivalent processes like ``paradoxical diffusion'' \cite{brockmann2006,magdziarz2007c,dybiec2009h}. In these cases, due to the competition between long waiting times ($\nu<1$) and L\'evy flights ($\alpha<2$), the second moment of the process $X(t)$ scales like $\langle x^2(t) \rangle\propto t^{2\nu/\alpha}$, i.e. for $2\nu=\alpha$ it assumes the form characteristic for a ``normal diffusion'', although $X(t)$ is non-Gaussian and non-Markov in nature.

The presence of L\'evy flights is visible in tails of the probability density functions. For $\alpha<2$,  the complementary cumulative distributions $F_c(x,t)$ demonstrate a power-law decay of the same type like $\alpha$-stable L\'evy densities governing distributions of jumps (see right panels of Figs.~\ref{fig:s09a09} -- \ref{fig:s05a10}).

In overall, numerical simulations corroborate that asymptotic (space) dependence of the process is determined by the jump length distribution while the rate of convergence to the long time asymptotics is determined by the subdiffusion parameter $\nu$, see bottom panels of Figs.~\ref{fig:s10a20} -- \ref{fig:s05a10}. This is particularly pronounced in the rate of decay of differences between theoretical and estimated probability densities ($M^2$ see Eq.~(\ref{eq:m2})) when for smaller value of the subdiffusion parameter the rate of convergence is significantly slower.

%
%
\section{Numerical issues\label{sec:numissues}}

The continuous time random walk with a one-parameter Mittag-Leffler distribution of waiting times and subordination method provide interesting and efficient frameworks for construction of solutions to the fractional Fokker-Planck equation (\ref{eq:ffpe}). Despite the fact that both methods can be used for solving the same fractional Fokker-Planck equation, there are some inherent differences among them which we want to discuss in more details.

\subsection{Precision}

Precision of the constructed results depends on the number of repetitions (MLF and subordination methods) and the time step of the integration (subordination method). Consequently, the method based on the subordination is more controllable, see bottom panels of Figs.~\ref{fig:s10a20} -- \ref{fig:s05a10}.

Figs.~\ref{fig:s10a20} -- \ref{fig:s05a10} compare analytical and numerical results. These figures indicate that the subordination method leads to higher level of agreement between approximate and exact solutions $p(x,t)$ of a corresponding fractional Fokker-Planck equation. First, subordination method reconstructs well not only the asymptotics (tails) of the solution but also its central part. Next, this method reconstructs the analytical solutions for all considered values of time $t$ while the CTRW scheme reproduces correctly the PDFs only in the asymptotic limits (i.e. after sufficiently long times  and for sufficiently large space excursions $x$). For $\nu\rightarrow 1$ the process $X(t)$ becomes a Markov L\'evy flight (see Fig.~\ref{fig:s10a10}). Its CTRW approximation scheme is then composed by use of the Mittag-Leffler function which for $\nu=1$ becomes an exponential distribution of waiting times. Also in this case, the convergence of both methods in reproducing PDFs $p(x,t)$ is met asymptotically.

\begin{figure}[!ht]
\begin{center}
\includegraphics[angle=0,width=8.0cm]{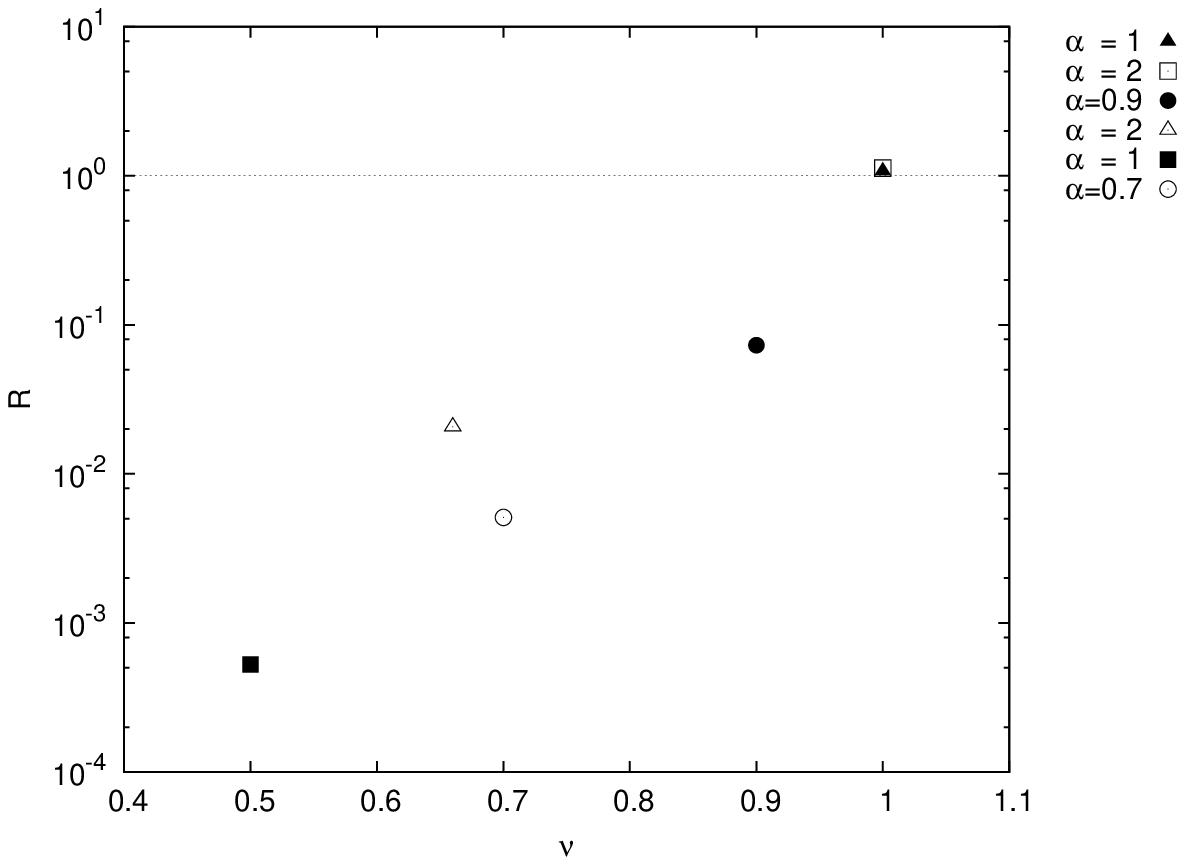}
\caption{Ratio $R$ between sums of squared differences $M^2$, see Eq.~(\ref{eq:m2}), for the subordination- and the CTRW- methods analyzed for different subdiffusion parameters $\nu$. Various symbols refer to different exponents $\alpha$. Data analyzed for $N=10^6$ trajectories at time $t=300$.}
\label{fig:errors}
\end{center}
\end{figure}

Finally, Fig.~\ref{fig:errors} quantify accuracy of both methods by comparing the ratio $R$ of sums of squared differences ($M^2$) for the subordination- and the CTRW- formalisms at time instant $t=300$. Direct analysis clearly indicates that the algorithm in which every path is generated as a subordination of two trajectories of the processes $\tilde{X}(s)$ and $S_{\nu}(t)$ leads to a higher precision in simulating adequate solutions to the fractional Fokker-Planck equation. The advantage of using the subordination technique is especially well visible for small values of the subdiffusion parameter $\nu$.

\subsection{Computational cost}

\begin{table}[!ht]
\begin{center}
\begin{tabular}{c|c||c|c||c|c|c|c}
\multicolumn{2}{c||}{} & \multicolumn{2}{|c||}{MLF} & \multicolumn{4}{|c}{subordination} \\
 \multicolumn{2}{c||}{$N$} & $10^5$ & $10^6$ & $10^5$ & $10^5$ & $10^6$ & $10^6$\\ \hline
\multicolumn{2}{c||}{$\Delta t$} & $-$ & $-$ & $10^{-1}$ & $10^{-2}$ & $10^{-1}$ & $10^{-2}$\\ \hline \hline

$\nu$ & $\alpha$ & \multicolumn{6}{c}{simulation time (seconds)} \\ \hline

1.0 & 2.0 & 1 & 17 & 7 & 75 & 76 & 763 \\ \hline
1.0 & 1.0 &1 & 14 & 4 & 48 & 46 & 457 \\ \hline
0.9 & 0.9 &1 & 13 & 12 & 118 & 122 & 1228 \\ \hline
0.7 & 0.7 &1 & 8 & 6 & 62 & 65 & 629 \\ \hline
2/3 & 2.0 &1 & 7 & 5 & 54 & 56 & 549 \\ \hline
0.5 & 1.0 &1 & 4 & 2 & 26 & 27 & 261 \\
\end{tabular}
\caption{Simulation time (in seconds) for various subdiffusion parameters $\nu$ and stability indices $\alpha$.
MLF -- relates to the CTRW framework with the Mittag-Leffler waiting time distribution, subordination -- to a subordination technique discussed in Section~\ref{sec:relation}. Results were averaged over $N$ realizations with the time step of the integration $\Delta t$. Number of repetitions is indicated in the 2nd row while the time step of the integration in the 3rd row. }
\label{tab:time}
\end{center}
\end{table}

The simulation time depends both on the number of repetition and the time step of the integration. The method based on the CTRW with waiting times distributed according to the Mittag-Leffler function is significantly faster but less precise. The performance of both methods is compared in Tab.~\ref{tab:time}. Exact values of the simulation time are provided for the informative purposes. The ratio between reported simulation times indicates which one of the methods is faster.

\subsection{Extensibility}

The subordination method is easily expendable to more general schemes of anomalous diffusion in external fields. In particular, the approach can be efficiently used to construct solutions to more general forms of fractional Fokker-Planck equations \cite{magdziarz2008,weron2008} with space and time dependent forces.
On the other hand,
the method based on the CTRW schemes is preferential when modeling anomalous diffusion as stemming from the underlying discretized version of a birth-and-death process \cite{heinsalu2006}.

Both methods can be easily extended to multidimensional cases. In such realms, conclusions drawn from one dimensional case still hold (numerical results not shown). However, some additional precaution is necessary when generating trajectories of multidimensional L\'evy flight or multidimensional continuous time random walks, as discussed elsewhere  \cite{teuerle2009} .

%
%
\section{Conclusions\label{sec:conclusions}}

Due to the asymptotic equivalence between various diffusion schemes and continuous time random walks the fractional Fokker-Planck equation plays a special role in statistical physics. The continuous time random walk scenario with power-law distributed waiting times and jumps lengths is asymptotically described by the fractional Fokker-Planck equation. In majority of situations it is not possible to solve this equation analytically. Very efficient and robust numerical approximations to its solutions are based on the stochastic representation of corresponding stochastic differential equations driven by stable noises \cite{janicki1994,nolan2010}.

In the paper we have examined asymptotic one dimensional diffusion stemming from two alternative approaches to CTRW.
The method based on the continuous time random walks with a one-parameter Mittag-Leffler distribution of waiting times provides an efficient way of construction of asymptotic solutions. Alternatively, the subordination method establishes tools which allow to reproduce PDFs of the fractional Fokker-Planck equation
within the whole time and space domains. Documented numerical accuracy and flexibility of the subordination method \cite{magdziarz2007,magdziarz2007c,koren2007c,dybiec-stationary}  results however in higher computational costs.

Altogether, the subordination technique provides a useful simulation tool \cite{magdziarz2007b} and clearly approximates  efficiently single time PDFs which solve the fractional Fokker-Planck equation. In contrast, the CTRW scheme with the Mittag-Leffler waiting time distribution reproduces correctly those PDFs only in the asymptotic limit.

Our analysis as presented in the paper relates to a free diffusion case, when the motion of particles can be described by the FFPE (\ref{eq:ffpe}), or alternatively rephrased in terms of the CTRW (or Langevin) description (Section~\ref{sec:relation}). This force-free case can be generalized to situations with inclusion of the inertia effect.
In this context, however, special care has to be taken when linking a particular subordination scheme with a type of the fractional Kramers-Fokker-Planck equation
\cite{magdziarz2007c,Heinsalu2007,eule2009}.
%
\begin{acknowledgments}

This project has been supported in part by Polish Ministry of Science and Higher Education (BD) and by Foundation for Polish Science (EGN) (International Ph. D. Projects Programme co-financed by the
European Regional Development Fund covering, under the
agreement No. MPD/2009/6; the Jagiellonian University International
Ph. D. Studies in Physics of Complex Systems).
\end{acknowledgments}

%
%

\begin{thebibliography}{10}

\bibitem{klages2008}
R. Klages, G. Radons, and I.~M. Sokolov, {\em Anomalous transport: Foundations
  and applications} (Wiley-VCH, Weinheim, 2008).

\bibitem{cox1965}
D.~R. Cox and H.~D. Miller, {\em The theory of stochastic processes} (Chapman
  and Hall, London, 1965).

\bibitem{taylor1998}
H.~M. Taylor and S. Karlin, {\em An introduction to stochastic modelling}
  (Academic Press, San Diego, 1998).

\bibitem{shlesinger1982}
M.~F. Shlesinger, J. Klafter, and Y.~M. Wong, J. Stat. Phys. {\bf 27},  499
  (1982).

\bibitem{meerschaert2006}
M.~M. Meerschaert and E. Scalas, Physica A {\bf 370},  114  (2006).

\bibitem{meerschaert2004b}
M.~M. Meerschaert and H.~P. Scheffler, J. Appl. Prob. {\bf 41},  623  (2004).

\bibitem{piryatinska2005}
A. Piryatinska, A. Saichev, and W. Woyczynski, Physica A {\bf 349},  375
  (2005).

\bibitem{saichev1997}
A.~I. Saichev and G.~M. Zaslavsky, Chaos {\bf 7},  753  (1997).

\bibitem{uchaikin2003}
V.~V. Uchaikin, Physics Uspekhi {\bf 46},  821  (2003).

\bibitem{tunaley1974}
J. Tunaley, J. Stat. Phys. {\bf 11},  397  (1974).

\bibitem{shlesinger1995}
{\em {L\'evy} flights and related topics in physics}, edited by M.~F.
  Shlesinger, G.~M. Zaslavsky, and J. Frisch (Springer Verlag, Berlin, 1995).

\bibitem{kotulski1995}
M. Kotulski, J. Stat. Phys. {\bf 81},  777  (1995).

\bibitem{nielsen2001}
{\em {L\'evy} processes: Theory and applications}, edited by O.~E.
  Barndorff-Nielsen, T. Mikosch, and S.~I. Resnick (Birkh\"auser, Boston,
  2001).

\bibitem{barkai2002}
E. Barkai, Chem. Phys. {\bf 284},  13  (2002).

\bibitem{srokowski2009}
T. Srokowski, Physica A {\bf 388},  1057  (2009).

\bibitem{metzler1999}
R. Metzler, E. Barkai, and J. Klafter, Europhys. Lett. {\bf 46},  431  (1999).

\bibitem{sokolov2000}
I.~M. Sokolov, Phys. Rev. E {\bf 63},  011104  (2000).

\bibitem{barkai2000}
E. Barkai and R.~J. Silbey, J. Phys. Chem. B {\bf 104},  3866  (2000).

\bibitem{chechkin2002c}
A.~V. Chechkin, R. Gorenflo, and I.~M. Sokolov, Phys. Rev. E {\bf 66},  046129
  (2002).

\bibitem{meerschaert2004}
M.~M. Meerschaert and C. Tadjeran, J. Comput. Appl. Math. {\bf 172},  65
  (2004).

\bibitem{gudowska2005}
E. Gudowska-Nowak, K. Bochenek, A. Jurlewicz, and K. Weron, Phys. Rev. E {\bf
  72},  061101  (2005).

\bibitem{szabat2007}
B. Szabat, K. Weron, and P. Hetman, Phys. Rev. E {\bf 75},  021114  (2007).

\bibitem{Heinsalu2007}
E. Heinsalu, M. Patriarca, I. Goychuk, and P. H\"anggi, Phys. Rev. Lett. {\bf
  99},  120602  (2007).

\bibitem{koren2007c}
T. Koren, J. Klafter, and M. Magdziarz, Phys. Rev. E {\bf 76},  031129  (2007).

\bibitem{sokolov2009}
I. Sokolov, E. Heinsalu, P. H{\"a}nggi, and I. Goychuk, EPL (Europhysics
  Letters) {\bf 86},  30009  (2009).

\bibitem{gudowskanowak2009b}
E. Gudowska-Nowak {\it et~al.}, Eur. Phys. J. E {\bf 30},  317  (2009).

\bibitem{baule2007}
A. Baule and R. Friedrich, Europhys. Lett. {\bf 79},  60004  (2007).

\bibitem{hilfer1995}
R. Hilfer and L. Anton, Phys. Rev. E {\bf 51},  R848  (1995).

\bibitem{fogedby1994}
H.~C. Fogedby, Phys Rev. E {\bf 50},  1657  (1994).

\bibitem{eule2009}
S. Eule and R. Friedrich, EPL (Europhys. Lett.) {\bf 86},  30008  (2009).

\bibitem{magdziarz2007}
M. Magdziarz, A. Weron, and K. Weron, Phys. Rev. E {\bf 75},  016708  (2007).

\bibitem{metzler2002}
R. Metzler and T.~F. Nonnenmacher, Chem. Phys. {\bf 284},  67  (2002).

\bibitem{podlubny1999}
I. Podlubny, {\em Fractional differential equations} (Academic Press, San
  Diego, 1999).

\bibitem{Chen2004}
Y. Chen, B.~M. Vinagre, and I. Podlubny, Nonlinear Dynamics {\bf 38},  155
  (2004).

\bibitem{gorenflo2007}
R. Gorenflo, F. Mainardi, and A. Vivoli, Chaos Solitons Fractals {\bf 34},  87
  (2007).

\bibitem{mura2008}
A. Mura and G. Pagnini, J. Phys. A: Math. Gen. {\bf 41},  285003  (2008).

\bibitem{scalas2006}
E. Scalas, Physica A {\bf 362},  225  (2006).

\bibitem{fulger2008}
D. Fulger, E. Scalas, and G. Germano, Phys. Rev. E {\bf 77},  021122  (2008).

\bibitem{magdziarz2007b}
M. Magdziarz and A. Weron, Phys. Rev. E {\bf 75},  056702  (2007).

\bibitem{magdziarz2008}
M. Magdziarz, A. Weron, and J. Klafter, Phys. Rev. Lett. {\bf 101},  210601
  (2008).

\bibitem{feller1968}
W. Feller, {\em An introduction to probability theory and its applications}
  (John Wiley, New York, 1968).

\bibitem{Note1}
Recall that the $\alpha $-stable motion $\left \protect \{L_{\alpha }(t),
  t\geqslant 0\right \protect \}$ is defined \cite {janicki1994} as a
  stochastic process of independent, stationary increments $\Delta L_{\alpha
  }=L_{\alpha }(t)-L_{\alpha }(t')$ distributed according to a stable law of
  index $\alpha $.

\bibitem{dybiec-anomalous}
B. Dybiec and E. Gudowska-Nowak, Phys. Rev. E {\bf 80},  061122  (2009).

\bibitem{dybiec2006}
B. Dybiec, E. Gudowska-Nowak, and P. H\"anggi, Phys. Rev. E {\bf 73},  046104
  (2006).

\bibitem{dubkov2008}
A.~A. Dubkov, B. Spagnolo, and V.~V. Uchaikin, Int. J. Bifurcation Chaos. Appl.
  Sci. Eng. {\bf 18},  2649  (2008).

\bibitem{kozubowski1999}
T.~J. Kozubowski and S.~T. Rachev, Int. J. Comput. Numer. Anal. Appl. {\bf 1},
  177  (1999).

\bibitem{magdziarz2007c}
M. Magdziarz and A. Weron, Phys. Rev. E {\bf 76},  066708  (2007).

\bibitem{rebenshtok2007}
A. Rebenshtok and E. Barkai, Phys. Rev. Lett. {\bf 99},  210601  (2007).

\bibitem{lomholt2007}
M.~A. Lomholt, I.~M. Zaid, and R. Metzler, Phys. Rev. Lett. {\bf 98},  200603
  (2007).

\bibitem{he2008}
Y. He, S. Burov, R. Metzler, and E. Barkai, Phys. Rev. Lett. {\bf 101},  058101
   (2008).

\bibitem{rebenshtok2008}
A. Rebenshtok and E. Barkai, J. Stat. Phys. {\bf 133},  565  (2008).

\bibitem{lubelski2008b}
A. Lubelski, I.~M. Sokolov, and J. Klafter, Phys. Rev. Lett. {\bf 100},  250602
   (2008).

\bibitem{dybiec2009b}
B. Dybiec, J. Stat. Mech.  P08025  (2009).

\bibitem{dybiec2009h}
B. Dybiec and E. Gudowska-Nowak, Phys. Rev. E {\bf 80},  061122  (2009).

\bibitem{metzler2000}
R. Metzler and J. Klafter, Phys. Rep. {\bf 339},  1  (2000).

\bibitem{sokolov2002}
I.~M. Sokolov, J. Klafter, and A. Blumen, Phys. Today {\bf 55},  48  (2002).

\bibitem{brockmann2006}
D. Brockmann, L. Hufnagel, and T. Geisel, Nature (London) {\bf 439},  462
  (2006).

\bibitem{weron2008}
A. Weron, M. Magdziarz, and K. Weron, Phys. Rev. E {\bf 77},  036704  (2008).

\bibitem{heinsalu2006}
E. Heinsalu {\it et~al.}, Phys. Rev. E {\bf 73},  046133  (2006).

\bibitem{teuerle2009}
M. Teuerle and A. Jurlewicz, Acta Phys. Pol. B {\bf 40},  1333  (2009).

\bibitem{janicki1994}
A. Janicki and A. Weron, {\em Simulation and chaotic behavior of
  $\alpha$-stable stochastic processes} (Marcel Dekker, New York, 1994).

\bibitem{nolan2010}
J.~P. Nolan, {\em Stable distributions - models for heavy tailed data}
  (Birkh\"auser, Boston, 2010), in progress, Chapter 1 online at
  http://academic2.american.edu/$\sim$jpnolan.

\bibitem{dybiec-stationary}
B. Dybiec, J. Stat. Mech.  P03019  (2010).

\end{thebibliography}

\end{document}